\documentstyle[12pt]{article}
\def\laq{\raise 0.4 ex \hbox{$<$}\kern -0.8 em\lower 0.62 ex\hbox{$\sim$}}
\def\gaq{\raise 0.4 ex \hbox{$>$}\kern -0.7 em\lower 0.62 ex\hbox{$\sim$}}
 
\def\beq{\begin{equation}}
\def\eeq{\end{equation}}
\def\bea{\begin{eqnarray}}
\def\eea{\end{eqnarray}}
\def\bq{\begin{quote}}
\def\eq{\end{quote}}

\def\frac#1#2{{\textstyle{{#1}\over {#2}}}}

\def\lsim{\mathrel{\rlap{\lower4pt\hbox{\hskip1pt$\sim$}}
    \raise1pt\hbox{$<$}}}
\def\gsim{\mathrel{\rlap{\lower4pt\hbox{\hskip1pt$\sim$}}
    \raise1pt\hbox{$>$}}}
\def\sqr#1#2{{\vcenter{\vbox{\hrule height.#2pt
         \hbox{\vrule width.#2pt height#1pt \kern#1pt
         \vrule width.#2pt}
         \hrule height.#2pt}}}}

\def\gappeq{\mathrel{\rlap {\raise.5ex\hbox{$>$}}
{\lower.5ex\hbox{$\sim$}}}}
 
\def\lappeq{\mathrel{\rlap{\raise.5ex\hbox{$<$}}
{\lower.5ex\hbox{$\sim$}}}}

\begin{document}

\begin{center}
{{\bf A note on the comments of arXiv:2003.10154 [gr-qc] to our paper Nonminimally Coupled Boltzmann Equation I: Foundations (arXiv:2002.08184 [gr-qc])}}

\vglue 0.5cm
{Orfeu Bertolami$^{1,2}$, Cl\'audio Gomes$^{2}$}

\bigskip
{$^{1}$ Departamento de F\'\i sica e Astronomia, Faculdade de Ci\^encias, Universidade do Porto\\}
\smallskip
{Rua do Campo Alegre 687, 4169-007 Porto, Portugal\\}

\vglue 0.3cm

{$^{2}$ Centro de F\'\i sica do Porto, Faculdade de Ci\^encias, Universidade do Porto \\}
\smallskip
{Rua do Campo Alegre 687, 4169-007 Porto, Portugal\\}
\end{center}

\vglue 0.7cm

\setlength{\baselineskip}{0.7cm}

\centerline{\bf  Abstract}
\vglue 0.7cm
\noindent
In this note we comment on remarks about our recent effort to understand the specificities of the Boltzmann equation in the context of the  alternative theories of gravity with non-minimal coupling between curvature and matter.

\vglue 1cm

\section{Introduction}

In a recent work, we have considered the Boltzmann equation \cite{nmcboltzmann} in the context of a class of alternative theories of gravity which extends f(R) theories by adding a non-minimal coupling between curvature and matter expressed through a product $f_2(R)\mathcal{L}$ in the action functional, for a new function of the scalar curvature $f_2(R)$ \cite{nmc}. By resorting to conservation laws arising from microscopic principles and aiming to match the macroscopic non-conversation of the energy-momentum tensor for a perfect fluid, typical of these gravity theories  \cite{nmc}, a condition on the normalisation of the distribution function was found. Given the mathematical formulation of the Liouville vector space discussed in Ref. [26] (O. Sarbach and T. Zannias, 2013) of Ref. \cite{nmcboltzmann}, it was found that the entropy vector flux is a non-decreasing function likewise in General Relativity (GR). Furthermore, a comment on the difference between Gibbs and Boltzmann formulations of the H-theorem was presented and implications for the model were drawn. We then analysed the case of an homogeneous and isotropic Universe and discussed the implications depending on the form of the function $f_2(R)$. Finally, we advanced some general remarks about the  gravothermal catastrophe in the context of this class of modified gravity theories. Of course, we are well aware that none of our considerations address the problems that one faces when attempting to show the H-Theorem in rigorous mathematical terms (see, for instance, Ref. \cite{Villani}).

Futhermore, we regard our work as a first attempt to study the deep connection between entropy and gravity in the  context of non-minimal curvature-matter coupling gravity theories (NMCCMGT), a connection which is known to exist, for instance, in black hole physics. And indeed, one of us (CG) has pursued the matter further on in a recent work \cite{CG_2020}. In this spirit, we were quite surprised to see the curious claim that the H-theorem does not hold in non-minimally coupled curvarture-matter models. In the note arXiv:2003.10154 [gr-qc], authors consider a single particle approach in the non-minimally coupled gravity and analyse a non-relativistic version of the collisionless Boltzmann equation to tackle relativistic considerations to surprisingly conclude that entropy variation arises from a collisionless Boltzmann equation! In what follows we shall discuss the misconceptions and misleading conclusions in each of the sections and sub-sections of the note arXiv:2003.10154:

\noindent
{\bf Nonminimally Coupled Gravity}

This section of the note arXiv:2003.1015 reproduces several discussions in the literature concerning the NMCCMGT proposed originally in Ref. \cite{nmc} assuming it might have a bearing on the discussion of Boltzmann's H-Theorem in the context of these theories. As mentioned the NMCCMGT arises from the Lagrangian density $f_1(R)+f_2(R)\mathcal{L}$, where, $f_1(R)$ and $f_2(R)$ are functions of the scalar curvature which, for phenomenological purposes, are expanded in a power series of $R$, modulated by a curvature, $R_0$, typical of the scale of the problem at hand. It follows that each term in the expansion has its own coefficient and is, in general, different from the others. The physical interpretation is that some terms of the expansion are suppressed by a physical mass scale. Obviously, the choice of the functions $f_1(R)$ and $f_2(R)$ must be compatible with GR where this theory provides a proper physical description. These assumptions are particularly suitable for addressing issues such as inflation, dark matter and dark energy, having no bearing on phases in the history of the Universe when it is dominated by radiation, which is known to be very well described by GR in the context of the Hot Big Bang model. 

\noindent
{\bf Force on Point Particles}

This sub-section copies, without quoting, the discussion of the original paper, Ref. \cite{nmc}.

\noindent
{\bf A note on the 4-acceleration of a fluid element}

This subsection is a repetition of the discussion of Ref. \cite{perfectfluids} where it is shown that the degeneracy of functional actions in what concerns the energy-momentum tensor of perfect fluids in GR, as discussed in the seminal papers of Refs. \cite{schutz,brown}, does not hold in the NMCCMGT. Even though this reference was not quoted, it is relevant to understand why the proposed choice of the note arXiv:2003.10154, namely ${\cal L} = T$ , where $\cal{L}$ is the Lagrangian density of matter and $T$ the trace of the energy-monentum tensor, is actually unsuitable.  

\noindent
{\bf Boltzmann's H-Theorem}

This section is particularly bizarre. First of all, the argument is based on Eq. (17) that is the non-relativistic collisionless Boltzmann equation. Furthermore, the authors of the note arXiv:2003.10154 seem no to understand that Boltzmann's H-Theorem concerns the deviation from the incompressibility of the distribution function in phase space due to the existence of collisions. Hence, it follows from the H-Theorem that the collisionless version of the Boltzmann equation implies that the distribution function is the equilibrium one concerning hence to adiabatic processes. The issue is well known and thorough discussions can be found in Refs. \cite{stewart,kremer}. The whole point of our paper, Ref. \cite{nmcboltzmann}, was to unsure that these conditions are respected in the gravity theory under study. 

\noindent
{\bf Entropy}

This section follows, without quoting, the remarks on our work \cite{nmcboltzmann} concerning Gibbs' and Boltzmann's H-theorem. 

\noindent
{\bf The Strength of Gravity}

The discussion presented in this section is completely besides the point. It assumes that $f_1(R)=R$ and drops the terms in the field equations that characterize the NMCCMGT. In fact, their considerations concern actually a sort of Rastall's theory of gravity \cite{Rastall}, which is known to be equivalent to GR \cite{Visser} and is not any NMCCMGT. Furthermore, the authors assume the contra-intuitive situation in which the gravitational interaction could be stronger at early and late times, which seems to hint for a collapsing Universe. In general, modified gravity models try to explain the large scale expansion of spacetime in its various phases and not its contraction given that there is evidence to consider this possibility. In any case, the pathologic choice of the note arXiv:2003.10154 [gr-qc] is not, by far, the one discussed in our paper. It should be pointed out that choices of the gravitational coupling should be considered having in mind implications they might have on the energy conditions as discussed in Ref. \cite{obms}

\section{Conclusions}

This section spells out clearly the bizarre reasoning of the note arXiv:2003.10154 [gr-qc]. Having assumed that the collisional term of the Boltzmann equation vanishes by not even considering it and by considering a non-covariant version of the collisionless Boltzmann equation, the authors claim to have found a problem in a type of Rastall's theory using entropy arguments allegedly due to Boltzmann's H-Theorem. This conclusion defies the well known conventional wisdom that, under collisionless assumptions, the H-Theorem can only imply that the distribution function is the equilibrium one and that space-time evolution is adiabatic. How can these misunderstandings have anything to do with the discussion developed in our paper \cite{nmcboltzmann} and lead to conclusions about NMCCMGT is a mystery that eludes any cartesian logics that is commonly used in physics.

We draw an end to this discussion stating that we have no intention to feed any polemics, but only to get the record straight in what concerns the physics discussed in our work \cite{nmcboltzmann} and the cumbersome comments expressed in the footnote arXiv:2003.10154 [gr-qc].  We welcome any discussion aimed to elucidate the matters at hand, but we shall not spent any further energy on matters that could have been clarified if addressed to us in a collegial way. If the usual professional procedures were followed, the footnote arXiv:2003.10154 [gr-qc] could putatively address some relevant aspects of Boltzmann's H-Theorem in the context of gravity theories beyond GR that were not covered in our paper. Rather unfortunately, the authors of the footnote arXiv:2003.10154 [gr-qc] did not follow this path and as such they incurred into a series of  mistakes and misunderstandings about beautiful and well known physics.

\end{document}